\documentclass[aps,prb,preprint,showpacs]{revtex4}
\usepackage{graphicx}
\usepackage{epsfig}
\usepackage{color}
\begin{document}
\topmargin-1.0cm

\title {
Tuning intermolecular interactions in di-octyl substituted
polyfluorene via hydrostatic pressure}

\author {K. Paudel, M. Chandrasekhar, and S. Guha} \email[Corresponding author E-mail:]{guhas@missouri.edu}
\affiliation {Department of Physics and Astronomy, University of Missouri-Columbia, Missouri 65211}

\date{\today}

\begin{abstract}
Polyfluorenes (PFs) represent a unique class of poly para-phenylene based blue-emitting polymers with intriguing structure-property
relationships. Slight variations in the choice of functionalizing side chains result in dramatic differences in the inter- and intra-chain
structures in PFs. We present photoluminescence (PL) and Raman scattering studies of bulk samples and thin films of dioctyl-substituted PF (PF8)
under hydrostatic pressure. The bulk sample was further thermally annealed at 1.9 GPa. The PL vibronics of the as-is sample red-shift at an
average rate of 26 meV/GPa. The thermally annealed sample is characterized by at least two phase transitions at 1.1 GPa and 4.2 GPa, each of
which has a different pressure coefficient for PL vibronics. The Huang-Rhys factor, a measure of the electron-phonon interaction, is found to
increase with increasing pressures signaling a higher geometric relaxation of the electronic states. The Raman peaks harden with increasing
pressures; the intra-ring C-C stretch frequency at 1600 cm$^{-1}$ has a pressure coefficient of 7.2 cm$^{-1}$/GPa and exhibits asymmetric line
shapes at higher pressures, characteristic of a strong electron-phonon interaction. The optical properties of PF8 under high pressure are
further contrasted with those of a branched side chain substituted PF.

\end{abstract}

 \maketitle
%******************************************************************************
\section{Introduction} \label{sec:intro}
Blue-emitting polymers continue to attract a lot of attention for
display applications. Among them polyfluorenes (PFs), which belong
to a class of poly para-phenylene (PPP) system, are extremely
attractive not only from the technological perspective due to their
high photoluminescence quantum yield, but also from the viewpoint of
serving as model systems for understanding fundamental photophysical
phenomena owing to their rich phase morphology. Molecular level
attributes such as local chain structure and side chain
conformations in these systems strongly impact their transport and
device characteristics. Slight variations in the choice of
functionalizing side chains result in dramatic differences in the
inter- and intra-chain structures in PFs. Highlighting these
differences are two prototypical PFs, poly (9,9-(di-octyl) fluorene)
(PF8) and poly (9,9-(di ethyl-hexyl) fluorene) (PF2/6). Side chain
substitution gives rise to a rich array of mesomorphic behavior with
the appearance of a nematic liquid crystalline (\emph{n}-LC) phase
at higher temperatures.\cite{grell} In the last decade many research
groups have focused on structural studies of PFs using x-ray
scattering,\cite{chen,knaapilaxrd,tanto} small angle neutron
scattering,\cite{rahman,knaapilaSANS} and optical spectroscopy.
\cite{cadby,winokur,liem,guha1}

PFs are characterized by 3-D crystalline phases as well as conformational phases or chromophores, where the latter depends on the structure of
the individual chains. PF2/6 has a limited number of conformational isomers that form a fivefold helix (5/2 or 5/1). \cite{lieser} These helices
in turn self-organize into three chain unit cells resulting in a well ordered semicrystalline hexagonal phase with coherence lengths exceeding
50 nm. \cite{tanto} Thus, the optical properties of PF2/6 are relatively insensitive to the exact crystallographic state, thermal history, or
molecular weight. PF8, on the other hand, has at least three conformational isomers that depend explicitly upon the torsional angle between the
adjacent monomers. \cite{chunwas} In addition to these isomers, structural studies of PF8 have identified many crystalline phases. \cite{chen2}

Recent studies have identified three different conformational isomers in terms of the local PF backbone torsional angle: C$_\alpha$ conformer
with a torsional angle of 135$^\circ$, C$_\gamma$ with an average torsional angle of 155$^\circ$, and the almost planar conformer, C$_\beta$,
has an average torsional angle of 165$^\circ$. \cite{chunwas} The C$_\beta$ conformer, conventionally knows as the $\beta$ phase has a separate
long wavelength feature in both optical absorption and emission spectra.\cite{cadby} Although this phase appears as a minority constituent, it
dominates the optical emission. Our recent work has shown that the morphology adopted by the side chains of PF8 is closely linked to the
structure of the molecular backbone. \cite{arif_PRL,volz} For example, the $\beta$ phase backbone is stabilized when the side chains in PF8
adopt an \emph{anti-gauche-gauche} conformation, as shown in the inset of Fig. 1. The rich phase morphology of PF8 is also seen in thin films of
its model oligomers.\cite{lidzey}

Hydrostatic pressure allows a study of materials in a region of phase space not accessible by chemical techniques. Application of hydrostatic
pressure allows tuning of both intermolecular and intramolecular interactions in polymer chains without changes in the chemical make-up.
\cite{guha2} The photophysical studies of conjugated materials under high pressure indicate that enhanced intermolecular interaction produces an
increased degree of conjugation, increased exciton trapping in organic polycrystalline molecules, \cite{wagner} redshift and broadening of the
photoluminescence, \cite{meera1,webster,gyang,morandiPPV} enhancement of excited state dynamics at polymer-polymer heterojunctions,
\cite{friendprl} destabilization of localized states as in methylated ladder-type PPP (MeLPPP), \cite{syang} and changes in the ring torsion
motion as in non-planar para hexaphenyl (PHP) and other oligophenyls. \cite{guha3,martin1} The pressure dependent optical studies in PHP and
other oligophenyls further reveal the individual contributions of the intermolecular and intramolecular interactions. \cite{guha3}

Optical studies in PFs under pressure allow tuning of the rich phase space and its impact on electron-phonon interactions. Our previous work on
PF2/6 under pressure shows dramatic changes in the PL spectrum that mainly originates from defect and aggregate states. \cite{martin2} The Raman
peaks shift to higher energies exhibiting unexpected antiresonance line shapes at higher pressures. Although the Raman spectra show very similar
behavior for PF8 and PF2/6 under pressure, indicating similar anharmonic interactions, the emission properties of the two materials are vastly
different under pressure. These differences arise as a result of their unique backbone conformations: PF2/6 due to its helical backbone
conformation shows a higher degree of overlap of the electronic wave-function, while PF8 also shows some excimeric emission but the changes in
its PL spectrum are much more gradual compared to PF2/6. Due to the various conformational isomers, PF8 shows planarization of the backbone at
relatively low pressures when pressure is applied to the more non-planar conformation, which is absent in PF2/6.

%*******************************************************************************
\section{Experimental Details}\label{sec:exptaldetails}
The PF8 sample was obtained from American Dye Source (BE-129) and loaded as-is in the pressure cell. For thin film studies under pressure, a
solution of PF8 (in toluene) was drop-casted on the bottom surface of the diamond. The sticking coefficient of PF films is quite poor on diamond
and therefore they experience a hydrostatic environment. The pressure studies were conducted in a Merrill-Bassett-type diamond anvil cell (DAC)
with cryogenically loaded argon as the pressure medium. Pressure was measured using the luminescence of a ruby chip located in the pressure
chamber. The photoluminescence (PL) spectra were excited using the 325 nm line of a HeCd laser or the 351 nm line of an Ar ion laser. The
luminescence excitation was analyzed with an Ocean Optics 2000 spectrometer with 25-micron slits. Raman spectra were collected using an Invia
Renishaw spectrometer attached to a confocal microscope with a $\times$50 long working distance objective and the 785 nm line of a diode laser
as the excitation wavelength. Typical laser power was a few mW on the sample. The bulk sample under pressure was also thermally cycled at room
temperature (RT) from its \emph{n}-LC phase at 160$^\circ$ C, using similar thermal steps as in previous ambient pressure studies.
\cite{arif_PRL}

%*******************************************************************************
\section{Steady-State Photoluminescence}\label{sec:Pl studies}
The PL studies were carried out from two samples of PF8 loaded in DACs, one from a thin film cast from toluene and the other from the bulk
sample. Both the film and bulk samples showed a signature of the $\beta$ phase before loading in the DAC. The bulk sample was thermally cycled
at 1.9 GPa by heating it to 160$^\circ$ C and bringing it slowly back to RT. Due to the higher self-absorption in the bulk sample, the high
energy vibronic peaks are not very well resolved. Vibronic progressions are clearly seen in thin films of PF8, indicating a coupling of the
backbone carbon-carbon stretch vibration to the electronic transitions. The transition highest in energy is the 0-0 transition, which takes
place between the zeroth vibronic level in the excited state and the zeroth vibronic level in the ground state. The 0-1 energy involves the
creation of one phonon.

The peak positions of the PL vibronics in PF8 are good indicators of
the nature of chain conformations; under ambient conditions the
$\beta$ conformer is identified by a 0-0 transition which is almost
$\sim$ 100 meV red shifted compared to the $\alpha$ conformer.
\cite{winokurAPL04} Figure 1 shows the PL spectra from the thin film
sample at selected values of pressure. The 2.83 eV 0-0 emission at
ambient pressure indicates the presence of the $\beta$ phase. The
Raman spectra from the film also indicate the presence of this
phase, as shown in Section IV. The PL vibronics are seen to red
shift upon enhanced pressures. The higher vibronic peaks smear out
beyond 5.0 GPa. The PL spectra in Fig. 1 were measured while
pressure was being increased. The hysteresis is small; upon
decreasing the pressure the PL spectra were almost identical to the
ones shown.

\begin{figure}
\unitlength1cm
\begin{picture}(4.0,8)
\put(-4,-1){ \epsfig{file=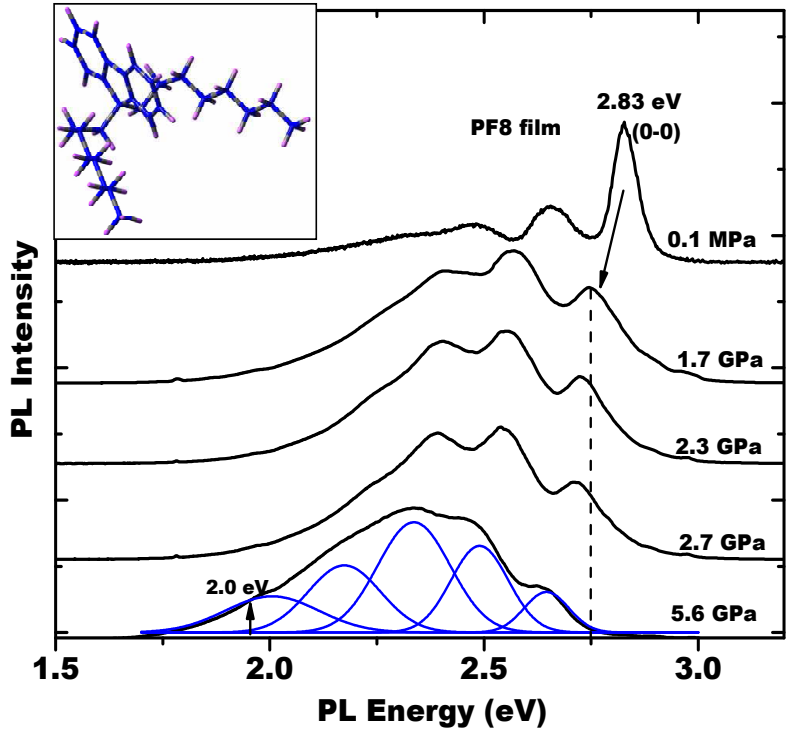, angle=0, width=12.cm, totalheight=10.7cm}}
\end{picture}
\caption{PL spectra of PF8 film at selected pressures measured at
room temperature. The inset shows a PF monomer with di-octyl side
chains in an \emph{anti-gauche-gauche} conformation.}
\label{figure1}

\end{figure}

Figure 2 compares the PL spectra of the as-is bulk sample before and after it was thermally cycled. The thermal cycling (TC) was carried out at
1.9 GPa, after which the PL spectra were measured upon increasing the pressure till 6.5 GPa, and then decreasing it down to 0.3 GPa. The ambient
pressure 0-0 PL vibronic peak blue shifts by almost 70 meV after TC. This is attributed to a change in the crystalline phase, discussed in
greater detail in the next section. This crystallization is seen even under pressure. The PL spectra further broaden after TC.

\begin{figure}
\unitlength1cm
\begin{picture}(4.0,7)
\put(-3,-0.5){ \epsfig{file=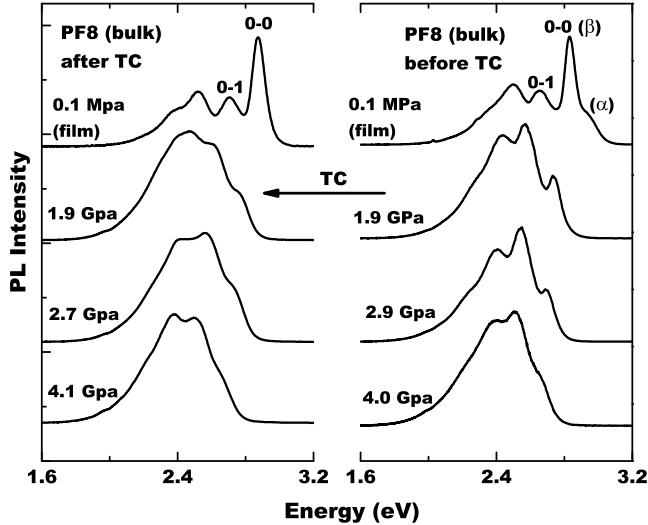, angle=0, width=10.cm, totalheight=8.7cm}}
\end{picture}
\caption{ The right panel shows the PL spectra of the bulk sample at selected values of pressure measured at room temperature. The left panel
shows the PL spectra of the same sample after undergoing thermal cycling. The 0.1 MPa data is from a thin film sample.} \label{figure2 }
\end{figure}

\subsection{Discussion of PL results}\label{sec:Pl results}
The complexity in deciphering the optical spectra of PF8 arises from the fact that the polymer must be simultaneously assessed in terms of its
crystallographic phase and the distribution of conformational isomers. PF8 cast from various solvents appear in a metastable structure at RT.
\cite{arif_PRL} Cooling from \emph{n}-LC phase yields the $\alpha$ and $\alpha'$ crystalline phase.  X-ray diffraction (XRD) studies show the
$\alpha'$ crystalline polymorph to be similar to the $\alpha$ phase and, in addition, it is exceptionally well oriented with respect to the
surface normal in both ultrathin and moderately thick films.\cite{chen} The progression to the $\alpha$ crystalline phase is extremely variable;
PF8 films cast from various solvents are marked by an intermediate \emph{M} phase at RT when cooled quickly. This phase is intermediate to
$\alpha$ and the $\alpha'$, and presumably corresponds to the C$_\gamma$ type family. A detailed Raman scattering study of PF8 as a function of
thermal cycling showed that heating the polymer to 130 $^\circ$C reduces the C$_\beta$ conformer and enhances the fraction of the C$_\alpha$
conformer. \cite{arif_PRL} Upon cooling the polymer back to RT from the \emph{n}-LC phase, a higher fraction of the C$_\gamma$ conformer is
seen.

Both the film and the bulk samples of PF8 in this work were initially characterized by the $C_\beta$ conformer. The red-shift of the PL
vibronics and the spectra as a whole, as seen in Figs. 1 and 2, are common to almost all conjugated polymers and molecules indicating a higher
degree of effective conjugation. This arises due to the higher overlap between the $\pi$-electron wavefunctions. The overall PL spectrum of PF8
were fitted with 4 Gaussian peaks, three of them originating from the 0-0, 0-1, 0-2 vibronics, and a low energy defect emission. Beyond 3 GPa
the PL spectra need to be fit with an additional low energy peak at 2.0 eV, as shown for the 5.6 GPa data in Fig. 1. Most probably this peak
arises from an excimer-type emission.

When PF8 is thermally cycled by cooling the sample slowly from its \emph{n}-LC phase, it changes to an overall three-dimensional \emph{M}
crystalline phase that precludes the $C_\beta$ conformer. \cite{volz} The resulting disordered $C_\alpha$ conformer, which has a non-planar
backbone conformation, shows a blue-shifted 0-0 peak at 2.9 eV (ambient pressure data in Fig. 2). Upon thermally cycling our PF8 sample at 1.9
GPa (indicated by the arrow in Fig. 2) the overall spectrum and the individual vibronics broaden. At this pressure the difference between the PL
vibronic energies between the thermally cycled and the as-is sample is almost 40 meV, similar to the trend seen in the ambient pressure data.
These results imply that even at 1.9 GPa the nature of the crystalline phase after TC is similar to thermal cycling at ambient pressure.

The peak positions of the 0-0 and the 0-1 PL vibronics are plotted in Fig. 3 for the film, and bulk sample before and after TC.  The open
symbols depict the 0-0 PL peak positions, and the filled symbols represent the 0-1 peak positions. The star symbols (filled and open) are the PL
peak positions from the as-is thin film sample that did not undergo any TC. The pressure coefficients of the 0-0 and 0-1 PL vibronic peaks are
listed in Table I. The pressure coefficient of the bulk (before TC) and the film samples are almost the same and yield an average pressure
coefficient of ($\sim$ -26 meV/GPa). The bulk sample after TC in contrast, shows three distinct regions with different pressure coefficients.
This is more pronounced for the 0-1 PL peak as shown by a linear fit of these regions (solid black line). The data suggests distinct phase
changes at 1.1 GPa and 4.2 GPa. Beyond 4.2 GPa the thermally cycled bulk sample shows a very slow red-shift of the PL energies. For comparison,
the pressure coefficient of the PL energies from a PF2/6 film are also shown. \cite{martin2} They are similar to as-is bulk PF8.

It is worth pointing out that in oligophenyls, where a single bond connects two phenyl rings, a sharp change in the pressure coefficient is
observed in the PL vibronics around $\sim$ 1.5 GPa; below 1.5 GPa the pressure coefficient of the PL energies is higher. \cite{guha3} Such
changes in oligophneyls have been attributed to planarization of the molecule. The repulsion between the ortho-hydrogen atoms leads to a torsion
of neighboring phenyl rings with respect to the single bond connecting them. First principles calculation of a PPP chain predicts a decrease in
the bandgap energy by almost 1 eV when the torsional angle between the phenyl rings changes from 50$^\circ$ to 0$^\circ$. \cite{Claudia1} Bulk
PF8 before TC behaves similar to PF2/6 \cite{martin2} with no discernable changes in the pressure coefficients as a function of pressure. Upon
TC, when the polymer crystallizes, the C$_\beta$ conformer is eliminated and the polymer has a high fraction of $C_\alpha$ or $C_\gamma$-type
chromophores. It is likely that at 1.1 GPa, pressure induces a planarization between the monomer units driving the system more towards the
C$_\beta$ conformation. Our Raman scattering studies presented in Section IV point in the same direction.

The 4.2 GPa discontinuity is not yet well understood. From our data it is clear that the sample is less compressible beyond 4.2 GPa in TC
samples. As of now the only XRD data under pressure that exist for the PF family (polymer/oligomer) is for crystalline fluorene, where a
reversible phase transition is seen at 3.6 GPa. \cite{heimel} This transition corresponds to a $\pi$ stacking molecular arrangement with an
increase in the bulk modulus (inverse of the compressibility).  The crystal structure of PF8 is quite different from fluorene itself but the
transition at 4.2 GPa in our data may be similar to fluorene with an enhanced $\pi$ stacking geometry, which would induce a stronger inetrchain
interaction. Additionally, theoretical calculations involving 3-D interactions of PPP chains show that an enhanced interchain interaction
results in a broadening of the absorption/PL with a very small red-shift of the band gap. \cite{syang} Our experimental data shows a large
increase in the PL linewidth of the TC sample beyond 5 GPa (Fig. 4), pointing to the direction that at these pressures the predominant effect of
hydrostatic pressure is an enhanced interchain interaction.

The PL linewdith shows differences between the as-is bulk and annealed samples. The full width at half maximum (FWHM) of the 0-1 PL vibronic is
plotted in Fig. 4; the 0-0 peak also shows a similar behavior. The as-is sample clearly shows a PL broadening with increasing pressures. PL
broadening as a function of pressure sheds light on the nature of intermolecular interactions and originates from the differences in
compressibility of the ground and excited electronic states in configuration coordinate space. \cite{drickamer} In amorphous polymers such as
MeLPPP, a similar broadening of the PL vibronics under pressure has been observed,\cite{guha2} reflecting a strong interaction perpendicular to
the chain axis. The PF8 sample that was thermally cycled, on the other hand, shows a very different dependence; on an average the PL vibronics
do not show any broadening till 6 GPa, beyond which a large increase in the FWHM is observed. No changes in the PL linewidths have been observed
in polycrystalline molecules such as PHP.\cite{guha2} Since TC induces an overall crystalline phase in PF8, no changes in the linewidths till 6
GPa indicate a lower compressibility perpendicular the chain axis. The sudden increase in the PL linewidths beyond 6 GPa implies a change in the
3-D crystalline phase. It is hard to predict exactly how the crystallinity changes beyond 6 GPa solely from optical measurements but these
observations match with the low pressure coefficient of the PL energies at high pressures (Fig. 3).

\begin{table}
\caption{Pressure coefficients for backbone PL emission peaks in PF8 bulk (before and after thermal cycling) and film for the 0-0 and 0-1 PL
peaks. The bulk sample after TC has been fitted in three regions. The pressure coefficients are determined by a linear fit to the PL energy
positions versus pressure given by $E(P)=E(0)+C P$. The last row shows the pressure coefficients of the PL vibronic energies in PF2/6.}

\label{table1}
\begin{ruledtabular}
\begin{tabular}{cccc}
Sample      & $C_{0-0}$(meV/GPa) &
  $C_{0-1}$ (meV/GPa)\\
\hline PF8 bulk (before TC)\   & -32.8$\pm$ 1.8 & -26.1$\pm$ 1.2\\
PF8 film (before TC)\  & -25.3 $\pm$ 0.9 & -21.6
$\pm$ 1.4\\
PF8 bulk (after TC) \ & -64.9 $\pm$ 6.5       & -90.8 $\pm$ 13.6\\
 & -48.3 $\pm$ 3.6       & -42.6 $\pm$ 1.5\\
  & -14.6 $\pm$ 5.3       & -4 $\pm$ 1.0\\
PF2/6 film\footnotemark[1]    & -40$\pm$ 1.0  & -28$\pm$ 1.0\\
\footnotetext[1]{Ref. \onlinecite{martin2}}
\end{tabular}
\end{ruledtabular}
\end{table}

\begin{figure}
\unitlength1cm
\begin{picture}(6.0,9)
\put(-4,-1){ \epsfig{file=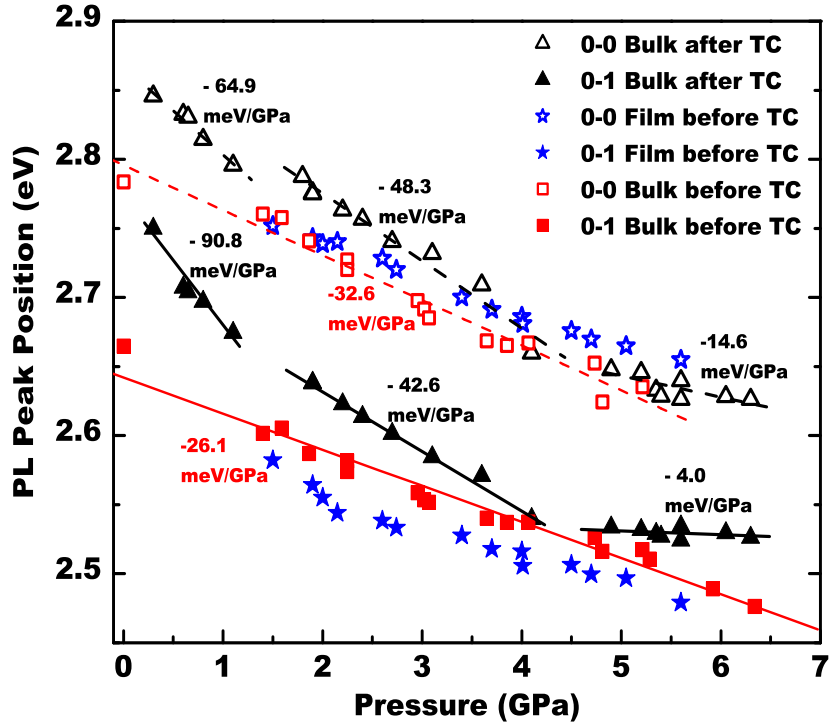, angle=0, width=13.cm, totalheight=11.7cm}}
\end{picture}
\caption{Peak positions of the 0-0 and 0-1 PL vibronics of PF8 under
pressure. The open and filled symbols represent the 0-0 and 0-1 peak
positions, respectively. The blue symbols denote the thin film
sample. The red and black symbols are from the bulk sample before
and after TC, respectively.} \label{figure3 }
\end{figure}

\begin{figure}
\unitlength1cm
\begin{picture}(4.0,7)
\put(-3,-0.5){ \epsfig{file=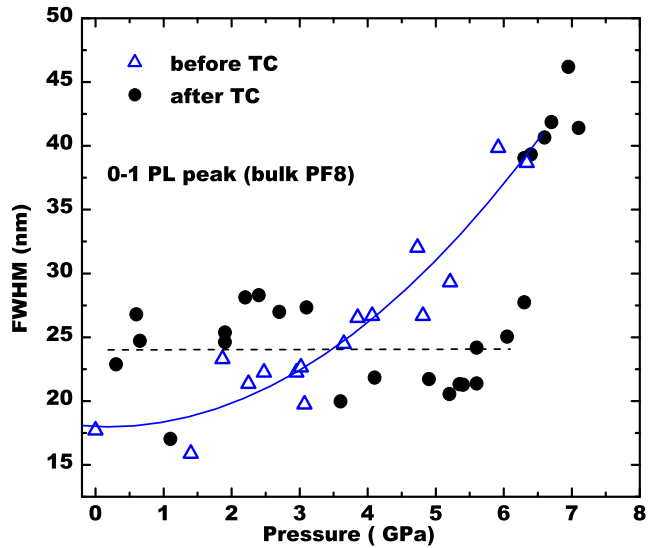, angle=0, width=10.cm, totalheight=8.7cm}}
\end{picture}
\caption{FWHM (in nm) of the 0-1 PL vibronic as a function of
pressure. The open $\triangle$ represents the as-is bulk sample and
$\bullet$ represents the bulk sample after TC. } \label{figure4 }
\end{figure}

\subsection{Pressure and temperature dependence of the Huang-Rhys factor}\label{sec:HR factor}
The fine features in the absorption and emission spectra of conjugated molecules/polymers are described by Frank-Condon (FC) coupling. The $\pi$
-$\pi*$ electronic transition is accompanied by a well resolved FC type progression of vibronic sub-bands. These vibronic bands are dominated by
modes representing local C-C stretching in the vicinity of 1200-1600 cm$^{-1}$. In the emission process, the Huang-Rhys factor (\emph{S})
corresponds to an average number of phonons that are involved when an excited molecule relaxes to its ground state configuration from its new
equilibrium configuration in the excited state (after the absorption of a phonon). Assuming that the vibrational frequency is the same for
ground and excited states and that the potentials are perfectly parabolic, \emph{S} may be experimentally determined from the fractional
intensity of the vibronic peaks. The relative intensities of the features coupled by a single phonon frequency ($\omega$) are described by

\begin{equation}\label{1}
\frac{I_{0\to n}}{I_{total}}=\frac{e^{-S}S{^n}}{n!},
\end{equation}
where $I_{total}$ is the total intensity of individual transitions. $I_{0\to n}$ is the intensity of the transition from the $0^{th}$ vibronic
excited state to the $n^{th}$ vibronic state of the electronic ground state. The Huang-Rhys factor therefore corresponds to the average number
of phonons that are involved when the excited molecule relaxes to its ground state configuration from the excited state, and $S\hbar\omega$ is
the relaxation energy. \emph{S} may be determined from the fractional intensity of the vibronic peaks:
\begin{equation}\label{2}
S=(I_{0\to 1}+2I_{0\to 2}+3I_{0\to 3}+...)/I_{total}.
\end{equation}
$I_{0\to 1}$, $I_{0\to 2}$, and $I_{0\to 3}$ refer to the intensity of the emission from the zeroth vibrational level excited state to the
first, second, and third vibrational level of the ground state, respectively.

The relative strengths of the vibronic transitions change with both
temperature and pressure. In a prior work we have shown that the
Huang-Rhys factor in small molecules and long-chain polymers
decreases with decreasing temperatures. \cite{guha1} The results for
PF8 as a function of temperature are shown in Fig. 5, where the
\emph{S}-factor was determined by using Eq. (2) and the vibronic
intensities beyond 0-3 were neglected. Smaller conjugated molecules
typically show a larger value for \emph{S}, which arises due to
their large normal coordinate displacement between the ground and
excited electronic states. The glassy phase in PF8 shows a much
higher value for the Huang-Rhys factor compared to the  $\beta$
phase, signaling planar conformations for the ground and excited
states of the latter. \cite{khan}

\begin{figure}
\unitlength1cm
\begin{picture}(4.0,7)
\put(-3,-0.5){ \epsfig{file=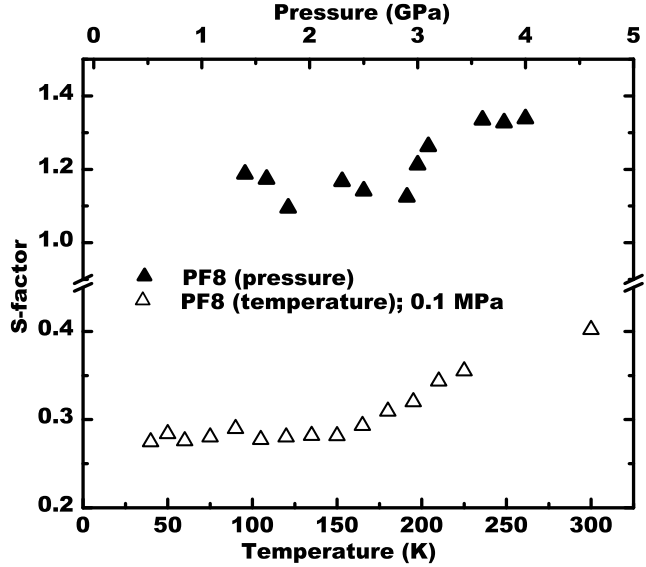, angle=0, width=10.cm, totalheight=8.7cm}}
\end{picture}
\caption{The Huang-Rhys factor versus temperature and pressure in
PF8.} \label{figure5 }
\end{figure}

At ambient pressure and RT the \emph{S}-factor in PF8 is determined as 0.4, similar to values obtained for the $\beta$-phase by Khan \emph{et
al}. \cite{khan} Figure 5 plots the \emph{S}-factor for PF8 both as function of temperature (at ambient pressure) and pressure (at RT) for a
thin as-is film sample. At 40 K the \emph{S}-factor is $\sim$ 0.25 increasing to $\sim$ 0.4 at RT. Upon increasing pressures, the
\emph{S}-factor is enhanced and almost remains a constant at 1.3, beyond 3.0 GPa. Lower values of the Huang-Rhys factor at ambient conditions
imply a delocalized excited state along with a smaller geometric relaxation that follows a transition from the excited state to the ground
state.

A comparison of the Huang-Rhys factors in PF8 as a function of temperature and pressure show that they increase, albeit at different rates. Upon
increasing the temperature, the singlet excitons typically become more localized in smaller conjugated chain segments. \cite{guha1} This
localization results in a higher value for \emph{S} with increasing temperatures. The red-shift of the PL spectra upon increasing pressures is a
clear sign of an enhanced effective conjugation and thus represents a higher delocalization of the excited state. This unambiguously shows that
a higher value of the \emph{S}-factor with enhanced pressures has a different origin as compared with enhanced temperatures; under enhanced
pressures the geometric relaxation of the electronic states is increased, increasing the \emph{S}-factor.

%*******************************************************************************
\section{Raman scattering studies}\label{sec:Raman}
The Raman spectrum of PFs is characterized by numerous
intramolecular C-C/C-H stretch and bend modes spanning from 100
cm$^{-1}$ to 1600 cm$^{-1}$. Both the vibrational frequencies and
intensities determined by Raman spectroscopy are strongly influenced
by variations in the backbone as well as side-chain conformations.
The Raman peaks in the 1600 cm$^{-1}$ region arise from an
intra-ring C-C stretch frequency and is best fit with two peaks: an
overwhelmingly dominant peak at 1605 cm$^{-1}$ and at least one or
two weak peaks in the range of 1570-1600 cm$^{-1}$. These weaker
peaks correspond to a breathing motion of the pentagon within the
monomer. The Raman frequencies in the 1250-1350 cm$^{-1}$ region are
associated with the backbone C-C stretch motion. Due to the strong
Raman peak of diamond, the 1200 cm$^{-1}$ region is not observed in
a DAC (Fig. 6 inset).

The low frequency Raman peaks in the 100-700 cm$^{-1}$ range, shown
in Figure 6, originate from the alkyl side chains. This region is
thus an extremely sensitive indicator of both side chain composition
and ordering. The ring torsion mode at 480 cm$^{-1}$ is seen in all
PFs with various side group substitutions, and is almost independent
of temperature. Since the Raman peaks around 600 cm$^{-1}$ and in
the 100-400 cm$^{-1}$ range are mainly from alkyl side chains, they
have very different signatures for various side chain substituted
PFs. There are predominantly two Raman peaks in the 600 cm$^{-1}$
region; the low frequency at 620 cm$^{-1}$ corresponds to a
stretching motion of the bridging C atom connected to the first
CH$_2$ group of the alkyl chain and the higher frequency peak at 633
cm$^{-1}$ involves a torsional motion of the phenyl rings. Shorter
alkyl chains such as ethyl-hexyl or an octyl group with
\emph{gauche} defects in PFs are characterized only by the higher
frequency Raman peak in this region. The relative ratio of the 620
and 633 cm$^{-1}$ Raman peaks in PF8 further track the presence of
the C$_\beta$ conformation. The intensity of the 633 cm$^{-1}$ peak
is stronger in the presence of the $\beta$ phase. It is clear from
our ambient pressure data in Fig. 6 that the PF8 film had a
significant fraction of C$_\beta$ conformation before loading the
DAC.

The 735 cm$^{-1}$ peak in PFs correspond to the 747 cm$^{-1}$
fundamental of a biphenyl with A$_1$ symmetry. The stronger Raman
peak in the 850 cm$^{-1}$ region is a combination of ring distortion
of the phenyl rings and a C-C stretch of the bridging carbon. The
weaker Raman peaks around 900 cm $^{-1}$ relate to C-C stretch type
motion of the alkyl side chains. \cite{volz} Frequencies in the
1100-1200 cm$^{-1}$ region (not shown in Fig. 6) are sensitive to
side-group substitution; the 1120 and 1135 cm$^{-1}$ peaks mainly
arise from the C-H bending modes (either local or between phenyl
units).

\begin{figure}
\unitlength1cm
\begin{picture}(4.0,7)
\put(-3,-0.5){ \epsfig{file=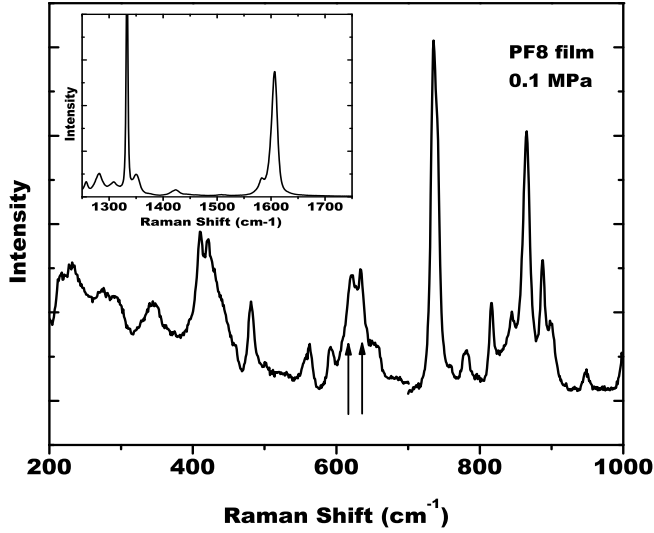, angle=0, width=10.cm, totalheight=8.7cm}}
\end{picture}
\caption{The Raman spectrum of as-is PF8 film at ambient pressure
and RT at low frequencies. The doublet at 600 cm$^{-1}$ marked by
the arrows track the conformational isomers. The inset shows the
high frequency region of the Raman spectrum of PF8 film; the sharp
peak at 1320 cm$^{-1}$ is the diamond peak from DAC.} \label{figure6
}
\end{figure}

\subsection{The 100-1200 cm$^{-1}$ region}\label{sec:low freq}

Figure 7 shows the ambient pressure data of a PF8 film sample before
and after TC along with the spectra of the bulk sample after TC
under pressure in the low frequency region. Due to the red-shift of
the PL spectrum under pressure, the background of the Raman spectrum
is quite high in this region. We select two pressure points which
are just above the discontinuity seen in the PL peak positions at
1.1 GPa. The ambient pressure results of PF8 after TC clearly shows
the appearance of new peak at 370 cm$^{-1}$. This is from the
longitudinal accordion motion (LAM) of all \emph{anti} conformation
of the alkyl side chain. Such a conformation of the side chains
preclude the $\beta$ phase suggesting that the individual backbone
conformation is C$_\gamma$ or C$_\alpha$-type. The two pressure data
upon TC show no signature of the 370 cm$^{-1}$ peak, which means
that the side chains at these pressures deviate from the all
\emph{anti} conformation, and the system is more like the C$_\beta$
conformation. This correlates well with the PL data, where a
different pressure coefficient is seen for the PL vibronics beyond
1.1 GPa, which we attribute to a more planar conformation of the
backbone. We note that in the presence of the $\beta$ phase, LAM
modes are not observed. Unfortunately, this low frequency region is
difficult to systematically track in Raman scattering for all values
of pressure due to the rising background.

\begin{figure}
\unitlength1cm
\begin{picture}(4.0,7)
\put(-3,-0.5){ \epsfig{file=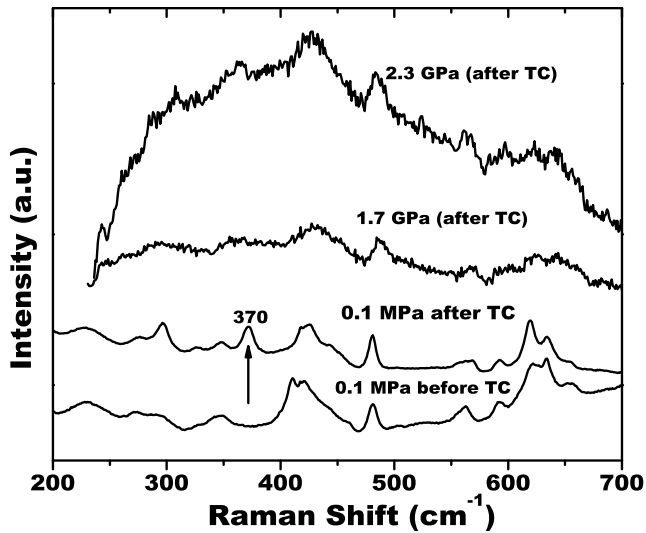, angle=0, width=10.cm, totalheight=8.7cm}}
\end{picture}
\caption{Low frequency Raman spectra of PF8. The bottom two spectra are at ambient pressure before and after TC. The top two spectra are at
elevated pressures after TC.} \label{figure7 }
\end{figure}

The Raman peaks in the 1140 cm$^{-1}$ region originate from a C–H in-plane bend motion along with a ring distortion. In the planar ($\beta$)
conformation of the polymer or in a monomer there is mainly one Raman band at 1135 cm$^{-1}$ that originates from the terminal phenyl rings. For
nonplanar conformations, this motion splits as two or more vibrations originating from monomer units about the center of symmetry of the
molecule and from the end rings. Upon thermal cycling the polymer from its \emph{n}-LC phase, which concomitantly induces an overall crystalline
phase and a reduction of the $\beta$ conformer, the 1135 and the 1172 cm$^{-1}$ peaks broaden as seen in the ambient pressure data in Fig. 8
(a). Under enhanced pressures there is a broadening of the 1135 cm $^{-1}$ peak; however, unlike ambient pressure, TC does not induce a further
broadening of this peak at high pressures. The pressure coefficient of the 1135 cm$^{-1}$ Raman peak is 1.6 cm$^{-1}$/GPa and 2.9 cm$^{-1}$/GPa
for the as-is and thermally-cycled samples, respectively. These pressure coefficients are lower compared to the high frequency 1600 cm$^{-1}$
Raman peak, as shown in the next section.

\begin{figure}
\unitlength1cm
\begin{picture}(4.0,7)
\put(-7,-1){ \epsfig{file=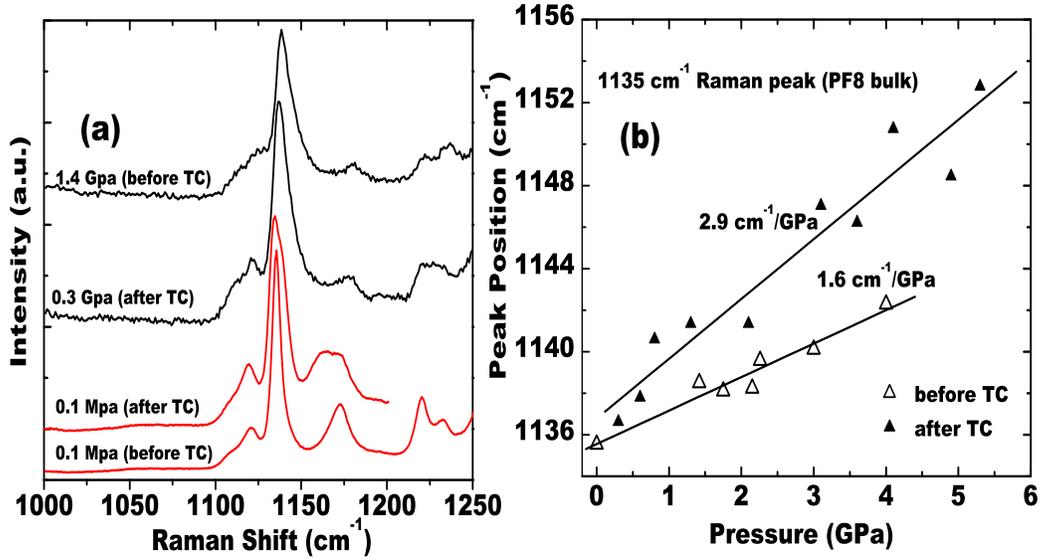, angle=0, width=17.cm, totalheight=10.7cm}}
\end{picture}
\caption{(a) The Raman spectra in the 1100 cm$^{-1}$ region at ambient pressure and higher pressures before and after TC. (b) Peak position of
the 1135 cm$^{-1}$ Raman peak before and after TC as a function of pressure. } \label{figure8 }
\end{figure}

\subsection{The 1600 cm$^{-1}$ region}\label{sec:high freq}

A signature of the various conformational isomers is better deciphered by the backbone C-C stretch modes in the 1300 cm$^{-1}$ region.
Unfortunately, under pressure in a DAC this region is swamped by the Raman peak of the diamond. The intra-ring C-C stretch peak at 1605
cm$^{-1}$ is clearly seen at all pressures for both as-is and TC samples. We present a detailed analysis of this peak as a function of pressure
in this section.

\begin{figure}
\unitlength1cm
\begin{picture}(4.0,7)
\put(-3,-0.5){ \epsfig{file=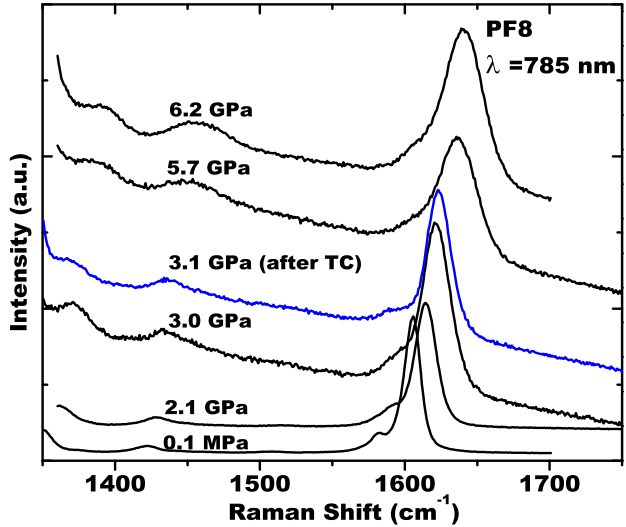, angle=0, width=10.cm, totalheight=8.7cm}}
\end{picture}
\caption{Raman spectra of the 1600 cm$^{-1}$ region (before TC) in PF8 at selected value of pressures. The spectrum at 3.1 GPa is from the PF8
sample after TC.} \label{figure9 }
\end{figure}

Since the origin of the 1600 cm$^{-1}$ Raman peak is from intra ring
C-C stretch motion, it is not sensitive to the various crystalline
phases or conformations. Fig. 9 shows a few selected spectra of the
1600 cm$^{-1}$ region of the as-is bulk sample under pressure; for a
comparison the 3.1 GPa spectrum of the TC sample is also shown.
Beyond 2.1 GPa the Raman background increases with a distinct
asymmetry of the 1605 cm$^{-1}$ peak, as shown in Fig. 9. Such an
asymmetry was observed in PF2/6 under pressure and was attributed to
a strong electron-phonon interaction between the Raman peaks and the
electronic continuum. \cite{martin2} This is characteristic of a
Breit Wigner Fano (BFW) resonance; in PF2/6 many of the vibrational
peaks also showed an anti-resonance behavior. To determine the peak
position, asymmetry parameter, and linewidth as a function of
pressure, we fit the 1605 cm$^{-1}$ peak with a BWF line shape given
by
\begin{equation}\label{3}
I(\omega)= I_{0} \: \frac{[(\omega - \omega_0)/q +
\Gamma]^2}{(\omega - \omega_0)^2 + \Gamma^2} \: ,
\end{equation}
where $\omega_{0}$ is the discrete phonon frequency, and $\Gamma$ is
the width of the resonant interference between the continuum and
discrete scattering channels. The asymmetry parameter $(1/q)$
depends on the average electron-phonon matrix element, $M$,  and the
Raman matrix elements between the ground and excited states of the
phonon and electron. The broadening parameter is given by
$\Gamma=\pi{M^2}D(\omega)$, where $D(\omega)$ is the combined
density of states for the electronic transitions.\cite{klein}

Figure 10 (a) shows the peak position of the 1605 cm$^{-1}$ Raman
peak as a function of pressure. The pressure coefficient of the bulk
PF8 sample before and after TC is almost the same at 7.2
cm$^{-1}$/GPa, which is higher than that observed for the 1135
cm$^{-1}$ Raman peak. Pressure-induced changes in the vibrational
frequencies is a measure of the Grun\"{e}isen parameter in the
potential energy surface of the particular vibrational coordinate.
For bulk inorganic solids the mode Grun\"{e}isen parameter is
usually volume-independent. In molecular solids due to the
differences between the local volume compression relative to that of
the bulk, a volume dependent Grun\"{e}isen parameter is often
observed. \cite{zallen} It is therefore not surprising that the
pressure coefficient of the 1135 cm$^{-1}$ and 1600 cm$^{-1}$ Raman
frequencies are different. We note that the mode Grun\"{e}isen
parameter, which is given by $(1/\omega)d\omega/dp$, is higher for
the 1600 cm$^{-1}$ peak compared to the 1135 cm$^{-1}$ peak.

The inset of Fig. 10 (a) shows the linewidth increasing almost
linearly with pressure for samples before and after TC. The
assymetry parameter (1/q) from a BWF fit is small below 2.0 GPa,
beyond which it increases rapidly with pressure as seen in Fig. 10
(b). Since the 1600 cm$^{-1}$ peaks originate from an intra-ring C-C
stretch motion, thermal cycling and thus the changes in the induced
phases have almost no impact on this Raman peak. The sign of the
asymmetry parameter in a BWF resonance, which arises from an
interaction of the electronic continuum, is an indicator of the
energy of the continuum. The negative value of q in the fits show
that the center of the electronic continuum lies below the discrete
mode frequency of 1605 cm$^{-1}$ (0.2 eV).

Since the vibrational frequencies of a harmonic solid are independent of compression, pressure induced changes in the Raman spectrum provide
insight into the anharmonicity of the solid state potential.\cite{ferraro} The linear shift of the 1605 cm$^{-1}$ Raman peak is a result of such
an anharmonicity of the potential. Although one expects an additional shift of the phonon frequencies due to the BWF interaction, it is not
possible to extract this information from our data due to the absence of a similar defect free polymer. The electronic continuum here arises
from defect states such as the fluorenone defects or excimeric states. We see subtle differences between PF8 and PF2/6; the latter showed a
higher asymmetry parameter and a square law dependence of the linewidth as a function of pressure. \cite{martin2} The optical properties of PF8
and PF2/6 under pressure are further contrasted in the next section.

\begin{figure}
\unitlength1cm
\begin{picture}(4.0,7)
\put(-7,-1){ \epsfig{file=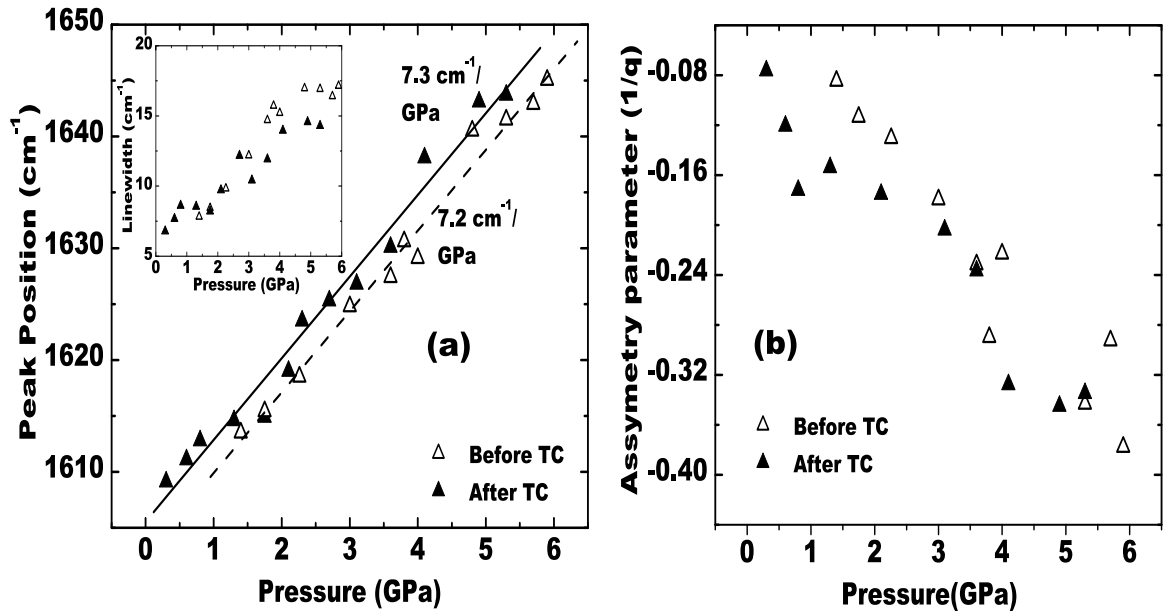, angle=0, width=17.cm, totalheight=9.7cm}}
\end{picture}
\caption{(a) Position of the 1605 cm$^{-1}$ Raman peak, obtained using a BWF fit, as a function of pressure for both as-is and TC PF8 bulk
sample. The inset shows the linewidth of same peak as a function of pressure. (b) Asymmetry parameter (1/q) of the 1605 cm$^{-1}$ Raman peak
versus pressure. 1/q is obtained by fits to the Raman peak with a BWF line shape [Eq. (3).]} \label{figure10 }
\end{figure}

%*******************************************************************************
\section{Summary and Prospect}\label{sec:conclusion}
A comparison of the luminescence studies of as-is and thermally cycled PF8 under pressure shows many differences in the crystallinity and
backbone conformations. Both samples show a red-shift of the PL energies upon increasing pressures but the TC sample shows three distinct region
with phase transitions at 1.1 GPa and 4.2 GPa. The 1.1 GPa is attributed to a planarization between adjacent monomers where the C$_\alpha$ and
C$_\gamma$ chormophores are driven towards the more planar C$_\beta$ form. The as-is PF8 sample under pressure shows a similar behavior to as-is
PF2/6 under pressure; the PL vibronics linearly shift with pressure although the nature of the emission at high pressures is quite different for
the two. Beyond 4 GPa PF2/6 shows an orange emission compared to the greenish emission seen in PF8 at higher pressures. The orange emission in
PF2/6 under pressure has been attributed to fluorenone defects and mixing of the singlet and charge-dipole states, which is enhanced due to the
helicity of the backbone.

The linewidth of the PL vibronics as a function of pressure shows
differences in the as-is and TC PF8 samples. The as-is sample shows
a broadening of the PL vibronics as a function of pressure. Such a
dependence is seen in other amorphous polymers and signals an
enhanced interchain interaction. The TC sample, which is in one of
the crystalline phases, shows almost no change in the PL linewidths
till 6 GPa. The main impact of pressure is to change the backbone
conformation of the polymer once the polymer is thermally cycled.
These results are similar to crystalline oligophenyls where there is
hardly any change of PL linewidths with increasing pressures.

Similar to PF2/6, the Raman frequencies in PF8 harden with pressure. Due to limitations in a DAC only the 1135 cm$^{-1}$ and the 1600 cm$^{-1}$
Raman peaks could be tracked as a function of pressure. These intramolecular vibrations are not the best candidates for unravelling changes in
the conformations or crystalline phases in PF8. The 1600 cm$^{-1}$ shows asymmetric line shapes characteristic of a BWF resonance, signaling a
strong electron-phonon interaction between the Raman phonons and the electronic continuum. The origin of the enhanced Raman linewdths is due to
such electron-phonon interactions.

In light of our experimental results on the optical properties of PF8 under pressure, where a rich change in the phase morphology is observed,
future XRD measurements under pressure would be invaluable to map these phase transitions in both PF8 and PF2/6.

%**********************************************************************************

\begin{acknowledgments}
This work was partly supported through the National Science
Foundation under grant No. ECCS-0523656. We thank M. Arif for the
ambient pressure Raman data.

\end{acknowledgments}
%***************************************************************************
%***************************************************************************

%***************************************************************************
%***************************************************************************
%***************************************************************************

\end{document}